\documentclass{pasj00}
\draft
\begin{document}
\title{Millimeter Light Curve with Abrupt Jump\\ in Cyg X-3 2008 April-May Outburst}
\SetRunningHead{T\sc{suboi} \rm et al. }{Cyg X-3}
\author{Masato T\sc{suboi}}
\affil{Institute of Space and Astronautical Science, Sagamihara, Kanagawa 252-5210}
\affil{Department of Astronomy, the University of Tokyo, Bunkyo, Tokyo 113-0033}

\author{Atsushi M\sc{iyazaki}}
\affil{Mizusawa VLBI Observatory, National Astronomical Observatory, Mizusawa, Oshu, Iwate 023-0861}

\author{Kouichiro N\sc{akanishi}}
\affil{Nobeyama Radio Observatory, National Astronomical Observatory, 
 Minamisaku, Nagano, 384-1305 }
\affil{National Astronomical Observatory of Japan, 2-21-1 Osawa, Mitaka, Tokyo 181-8588}
\author{Taro K\sc{otani}}
\affil{Department of Physics and Mathematics, Aoyama Gakuin University, Sagamihara, Kanagawa 252-5258 }

\KeyWords{}
\maketitle

\begin{abstract}
Cyg X-3 is a well-known microquasar with a bipolar relativistic jet.  Its
famous giant radio outbursts have been repeated once every several years. However, the  behavior of the millimeter wave emission has remained unclear because of limitations of time resolution 
 in previous observations. We report here millimeter wave observations
of Cyg X-3 experiencing giant outbursts with one of the finest time resolutions. 
We find  a series of short-lived flares with amplitude of 1-2 Jy in
the millimeter light curve of the 2008 April-May outburst. They have flat spectra around 100 GHz.  We also find abrupt and large amplitude flux density changes with $e$-folding time of 3.6 minutes or less. 
The source size of Cyg X-3 is constrained within $0.4$ AU and the brightness
temperature is estimated to be $T_\mathrm{B}\gtrsim 1\times10^{11}$ K. 
\end{abstract}
 
\section{Introduction} 
Cyg X-3 is a famous microquasar containing a
Wolf-Rayet star and a compact object, whose nature or mass is not
known (\cite{Lomment2005}). Its giant radio outbursts have been observed once every several
years.  The centimeter emission of Cyg X-3 during the bursts is composed
of superposition of a series of short-lived flares and relatively static
low-level emission (e.g. \cite{Ogley2001}).  VLBI observations revealed that Cyg X-3 has sometimes a 
bipolar relativistic jet  (\cite{Miller-Jones2004},  \cite{Tudose}) and sometimes  a one sided jet (\cite{Mioduszewski}).
This is presumably the remnant of ejecta by the flares.  Although the
radio emission arises through synchrotron process of relativistic
electrons in the jet (\cite{Hjellming1988}, \cite{Miller-Jones2004}),  the behavior at millimeter-wave during the flares is not yet established.
During the outbursts of 2006, we found that time scale of
the flares shortens apparently at the inverse of the frequency (Tsuboi
et al. 2008).  A millimeter wave observation with a higher time
resolution is desirable to understand the mechanism of the flares. Then
it was reported that spectacular large outburst of Cyg X-3 occurred
again at April 2008 (\cite{Trushkin2008}).

\section{Observations}
 We monitored the flux density of Cyg X-3 with a high time resolution of 4-6 min using Nobeyama Millimeter Array (NMA) of  Nobeyama Radio Observatory (NRO)\footnote{Nobeyama Radio Observatory is a branch of National Astronomical Observatory, National Institutes of Natural Sciences, Japan} for four days in the period from MJD$= 54592.85$ (May 7 2008) to $54597.97$ (May 12 2008) at both 90 and 102 GHz simultaneously.  The NMA consists of six 10-m element antennas equipped with double side band (DSB) superconductor-insulator-superconductor (SIS) receivers with a single linear polarization feed.  We used four element antennas, B, C, D, and E in the first three days and  five element antennas, B, C, D, E, and F in the last day. The Ultra-Wide-Band Correlator (UWBC) with a 1GHz bandwidth was employed for the backend (Okumura et al. 2000).  The quasar J2015+371 was observed every 17 minutes as a phase and amplitude reference calibrator. Neptune was used as a primary flux-scale calibrator.  The observed flux densities of J2015+371 at 90 and 102 GHz were both 3.3 Jy during the four days.  

Fig.1a and Fig.1b show the examples of the UWBC output in the second day. They show the visibility amplitudes of Cyg X-3 and J2015+371 on the baselines between antennas B-E and C-D, respectively.  These are independent except for atmosphere variation.  Fig.1c and Fig.1d show the gain variabilities of the baselines derived from the visibility amplitudes of J2015+371. 
The UV-data were calibrated with the UVPROC-II software package developed at NRO (Tsutsumi et al. 1997).  To obtain the flux densities averaged every 4-6 min for a short timescale light curve, we applied a point source model to the visibility data self-calibrated for phase using the ATNF MIRIAD package. We removed low quality data at low $EL$ to get final results. For example, we used only the data from MJD$=54593.85$ to $54594.03$ to make figure 4b.

\clearpage
\begin{figure}
\begin{center}
\includegraphics[width=16cm]{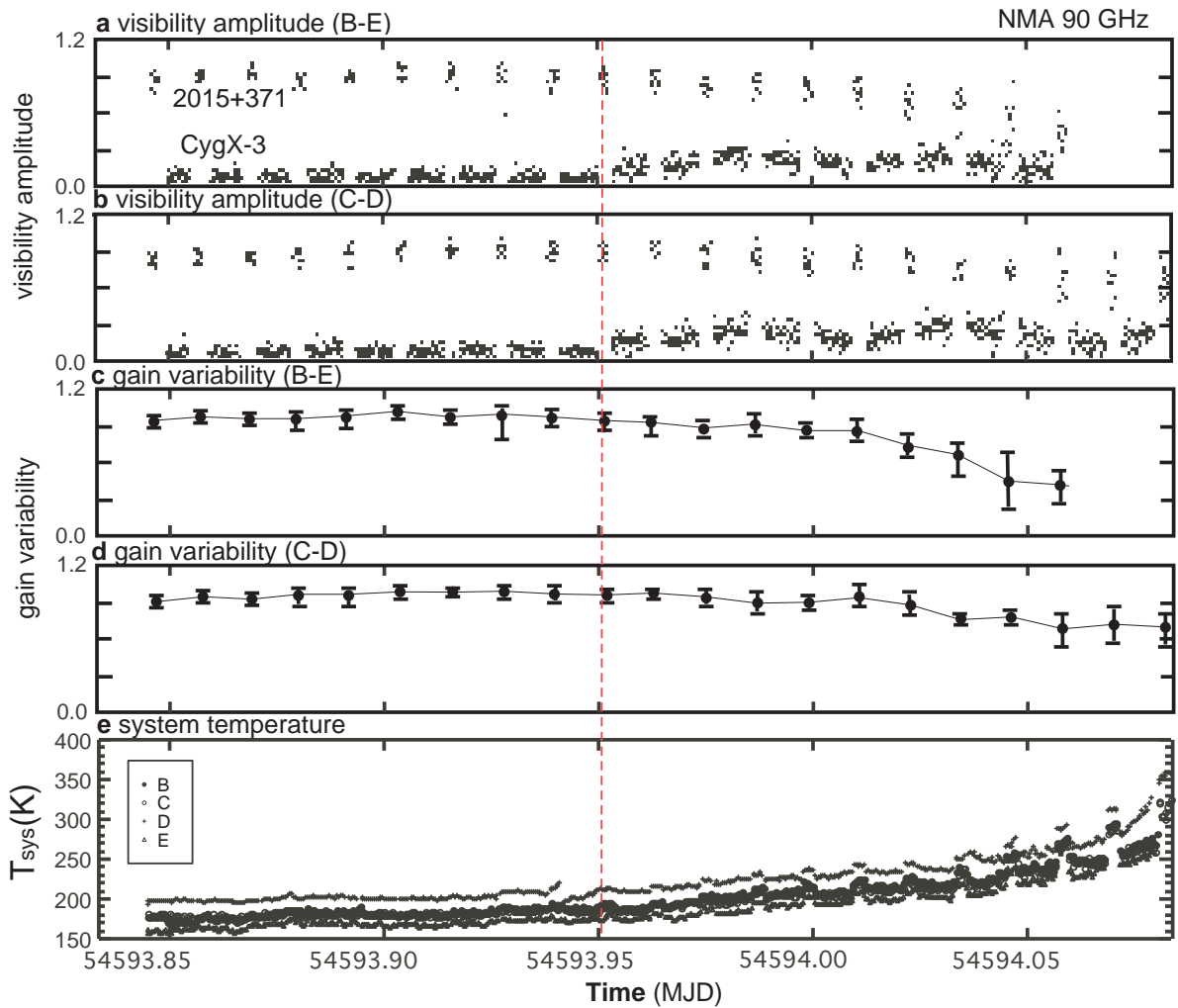}
\caption{{\bf a:} Visibility amplitudes of Cyg X-3 and the flux calibrator, J2051+371 on the baseline of antennas B and E of the NMA in MJD$=54594$. The frequency band is 90 GHz. The red vertical broken line indicates the timing of the intensity jump of Cyg X-3 shown in Fig. 4b. {\bf b:}  Visibility amplitudes at 90 GHz on the baseline of antennas C and D  of the NMA. {\bf c:}  Gain variability on the baseline of antennas B and E derived from the visibility amplitude of the flux calibrator, J2051+371.  {\bf d:}  Gain variability on the baseline of antennas C and D. {\bf e:}  Variability of system temperatures of the element antennas B, C, D, and E. }
\label{Fig1}
\end{center}
\end{figure}
\clearpage
 
Fig.1e shows the variability of the system noise temperatures including atmospheric attenuation and antenna ohmic loss of the element antenna of the NMA. The system noise temperatures were less than 250 K in both frequency bands during the observations of the four days.
The difference of atmospheric attenuation between Cyg X-3 and J2015+371 may produce artificial change of the flux density of Cyg X-3. We monitored the optical thickness of atmosphere through the system noise temperature. The thickness gradually change in the range of $\tau =0.1-0.4$ during the observations. The monitored thickness was applied for the correction of the flux densities. The difference of atmospheric attenuation may cause calibration error up to 10 \% around lowest elevation, $EL \sim 30^{\circ}$. The total systematic uncertainty of absolute flux density is at most 15 \%. The remaining error is mainly caused by the calibration error between the primary flux-scale calibrators and  J2015+371, and phase noise by atmospheric fluctuations. Therefore, the relative uncertainties of the flux density for adjacent bins should be better than the value.

\section{Millimeter Light Curve of Cyg X-3}

The flux densities of Cyg X-3 at 90 and 102 GHz are plotted as a
function of time in Fig.2. The first bin of our observation is at
MJD$=54592.84$ (May 7 2008), about 19 days after the first rise detected
with RATAN-600 in April 2008 (\cite{Trushkin2008}).  
There are the series of short-lived flares with amplitude of $1-2$ Jy in MJD$= 54593$ (May 7 2008). Although there are daily intermissions between individual observations because
Cyg X-3 is not a circumpolar star at Nobeyama,  the peak flux density of the flares, or the activity of the outburst,  gradually decreases with time in three days.  
Eventually, the flux densities at MJD=54598, 12 May 2008, decrease to level of 100 mJy, which is similar to those of the non-burst phase.  Assuming the exponential decay of the
activity, the e-folding decay time is estimated to be about 1.25 day, or 30 hours.  

Fig.3 shows the spectral index, $\alpha$ ($S_\nu=S_{\nu 0}\nu^{-\alpha}$),  between 90 and 102 GHz as a function of the average flux density at 100 GHz. The spectral indexes of  the flares with over 1 Jy are almost zero. The average spectral index is $\alpha(>1\mathrm{Jy})= -0.06\pm0.26$. This shows that the flares in MJD$= 54593$ (May 7 2008) have flat spectra. 
 In addition, the flares with 0.5-1 Jy have flat spectra. 
The spectra suggest that the flares have medium optical thickness, or that the mm-wave synchrotron emitting electrons are being accelerated so that 
they have the energy spectrum of $N(E)\propto E^{-1}$. 
On the other hand, the average spectral index of low level emission is $\alpha(<0.4 \mathrm{Jy})= -0.88\pm1.04$. The spectral index may be inverted although the scatter of the data is fairly large.   
The inverted spectrum may suggest optically thick synchrotron emission from a partially self-absorbed jet 
because optical thickness can vary the spectral index from $\alpha= 0$ down to $\alpha= -2.5$.  

\begin{figure}
\begin{center}
\includegraphics[width=15cm]{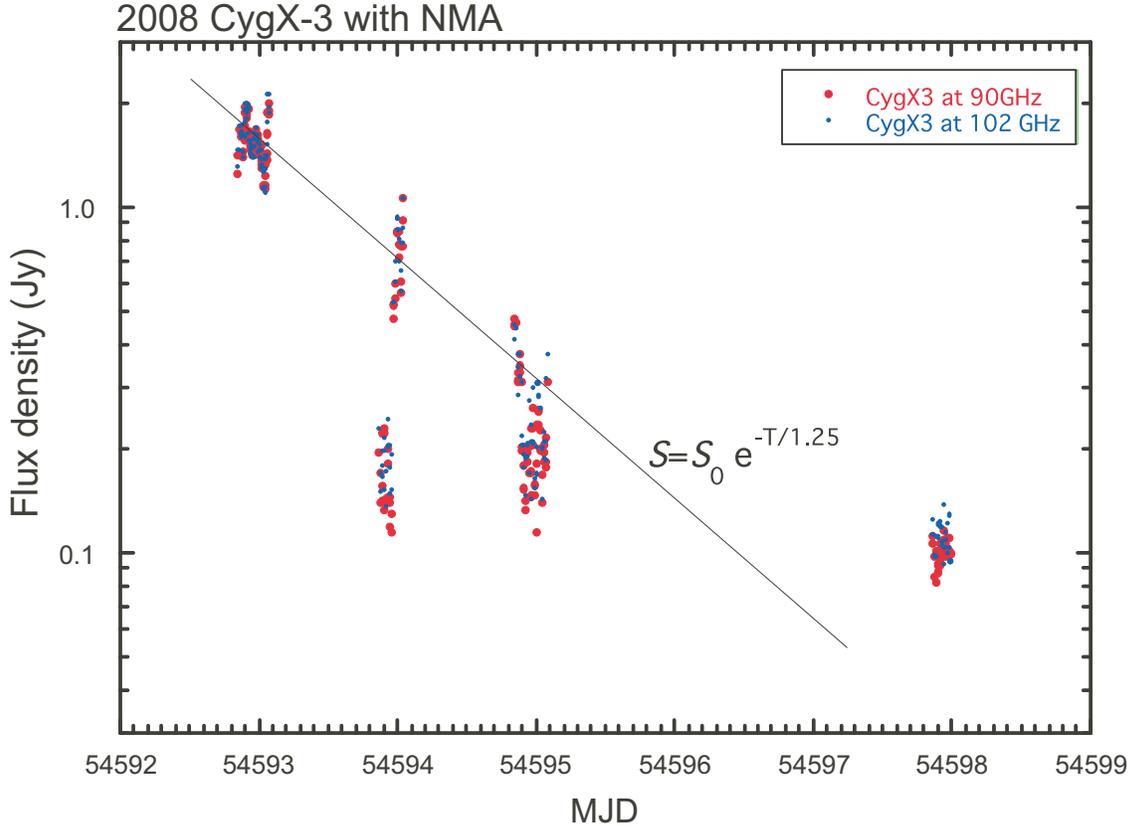}
\caption{The light curves of Cyg X-3 with NMA at $90$ (red circles) and $102$ (blue circles) GHz from MJD$= 54593$(7 May 2008) to MJD=$54598$(12 May 2008).  The first epoch is about 19 days after the first rise  at MJD$=54574.16$(18 April 2008)(\cite{Trushkin2008}).  Although there are the series of short-lived flares with $1-2$ Jy in MJD$= 54593$(7 May 2008) the peak flux density of the flare decreased for three days.  The flux densities at MJD$=54598$(12 May 2008) deceased to level of $100$ mJy. {\bf The solid line express an exponential decay model with an e-folding time of 1.25 d, or 30 h.}}
\label{Fig2}
\end{center}
\end{figure}
\begin{figure}
\begin{center}
\includegraphics[width=16cm ]{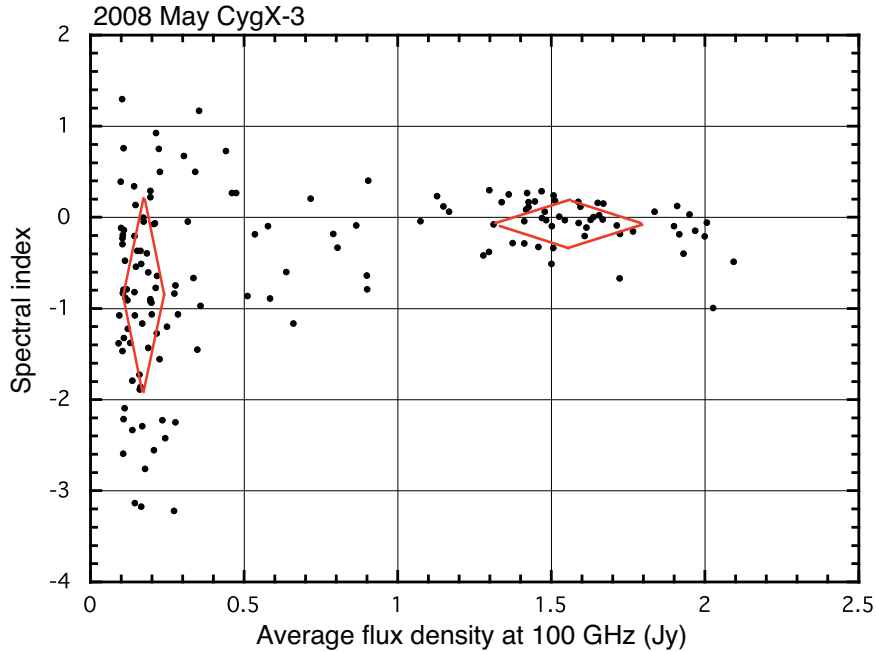}
\caption{Spectral index, $\alpha$ ($S_\nu=S_{\nu, 0}\nu^{-\alpha}$),  between $\nu=90$ and $102$ GHz is plotted as a function of the average flux density. There may be two states with the soft higher flux density and the hard lower flux density. The average spectral indexes are $\alpha= -0.06\pm0.26$ for high level emission $(S_\nu > 1\mathrm{Jy})$ and $\alpha = -0.88\pm1.04$ for low level emission $(S_\nu < 0.4 \mathrm{Jy})$, respectively. Red open diamonds show the average values and errors. }
\label{Fig3}
\end{center}
\end{figure}

\section{ Abrupt Intensity Jump of Cyg X-3} 
 Fig. 4a--d show the mm-wave daily light curves of Cyg X-3 with a higher
time resolution.   In Fig.~4-a, the source undergoes successive flares so that the
flux densities does not decay bellow 1~Jy.  The observed peak flux densities at 90 and 102 GHz are up to 2.2 Jy. 
The $e$-folding rise and decay time of the peaks is estimated to be $\sim 1$ h or shorter, although the precise times of the beginning and end of variations are difficult to determine.
Flux densities at 102~GHz (blue dots) are slightly larger than those at 90~GHz (red dots) in the rising phase of the peaks, and slightly less in the decay phase. This characteristic suggests that the change at 90 GHz is somewhat delayed compared with that at 102 GHz  in the both phases. The delay, if any, are no more than a few minutes.

In Fig. 4-b, we observed a few peaks exceeding 1 Jy. There is a large jump in flux density at MJD=54593.95 (a black arrow).  Within 5 minute, the flux densities increase four times or more compared with the value of the previous scan. 
The sudden increase at MJD=54593.95 in visibility amplitudes of Cyg X-3 is also identified in Fig.1a and 1b.  Because this was observed on two independent baselines, an antenna-based error, such like pointing-error, can not account for the observed jump.  
An artificial intensity jump due to abrupt change in the visibility amplitude of J2015+371, baseline gain variability, or system temperatures, is unlikely, because no such change is identified in these data of Fig. 1. 
In addition, the uncertainty caused by the difference of atmospheric attenuation was at most  several \% around the elevation of the observed jump. Then we conclude that the abrupt intensity increase of Cyg X-3 must be a real event. 
Time delay between 90 and 102 GHz is not observed in these data. The abrupt increase may be explained in terms of the abrupt injection of synchrotron emitting electrons.  The $e$-folding time of the increase is
about $t_e=3.6$ minute. This is much shorter than the $e$-folding rise time observed in Fig. 4-a.
Physical diameter of the mm-wave emitting region, $2r$, is estimated
to be less than $2r\le c\times t_e= 6.5\times 10^{10}$ m $= 0.43$ AU, from light crossing in the $e$-folding time. 
 There are two flares with peak flux densities of 1~Jy in the
later half of the light curve shown in Fig.~4b (MJD=54594). The changes in visibility amplitudes of Cyg X-3 are also visible in Fig.1a and 1b. Such change is not identified in gain and system noise (see Fig 1c, 1d, and 1e). The amplitudes of the change at both frequencies are as large as $\frac{\Delta S}{S_{min}} =100$ \%. Time delay of the change between 90 and 102 GHz is not observed in the data. The $e$-folding times of the flares are
less than $t_e \le 60$ minutes. These are similar to the series of flares observed in MJD$= 54593$.  

There may be another sudden change at MJD=54594.88 (a black arrow) in Fig 4c. Within 5 minute, the flux density of Cyg X-3 decreases about half although the visibility amplitude of J2015+371, the baseline gain variability, and the system temperatures were not abruptly changed. This is similar but opposite-sense to the jump observed at MJD=54593.95. 

\begin{figure}
\begin{center}
\includegraphics[width=14cm]{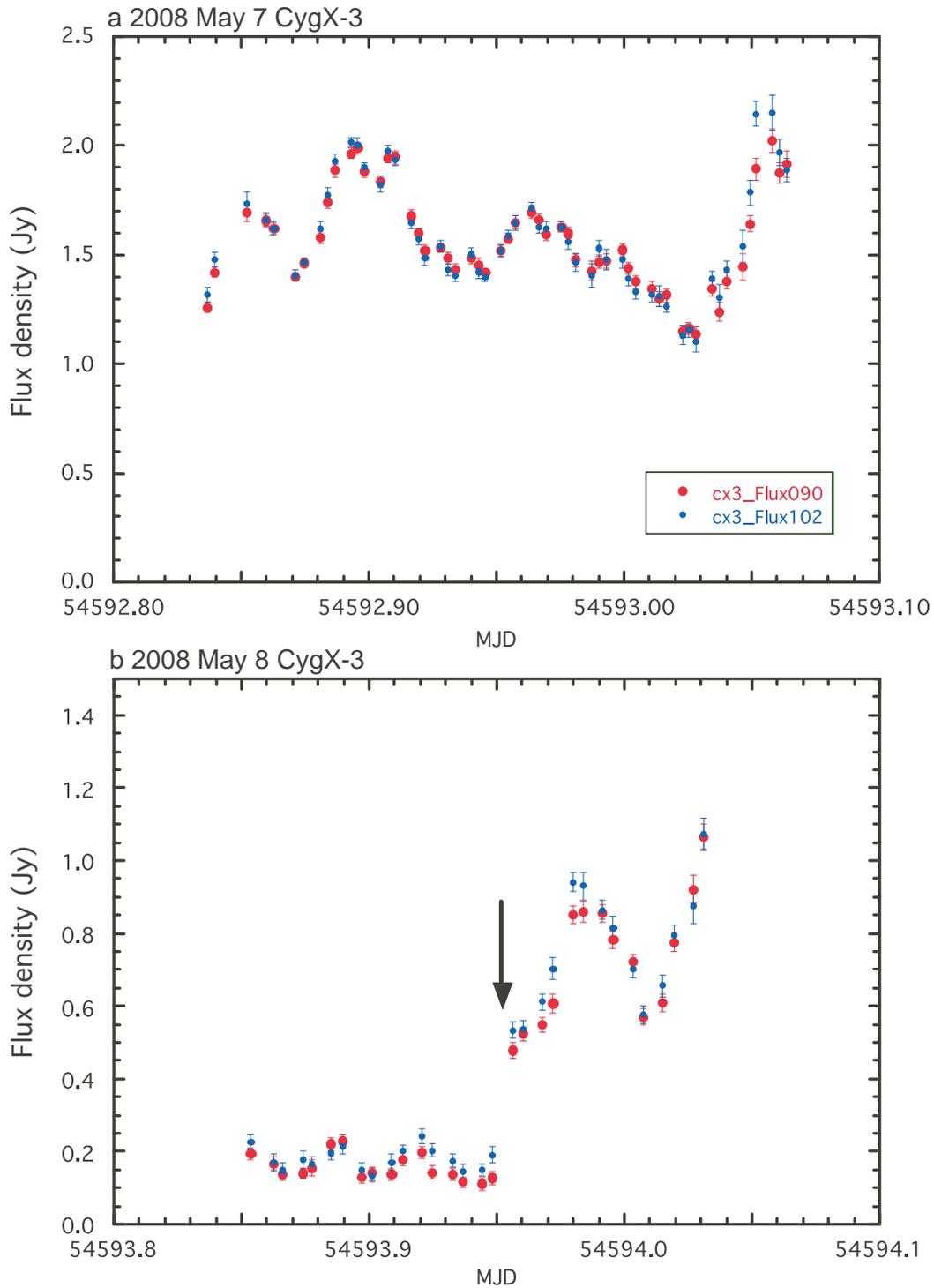}
\caption{\bf a: \rm The mm-wave light curves of Cyg X-3 with NMA at $90$ (red filled circles) and $102$ (blue filled circles) GHz from MJD$=54592.84$ to $54593.07$. There are several short-lived flares with duration of less than one hour.  The observed peak flux densities at $90$ and $102$ GHz were about $2.2$ Jy.  The ratio of these flux densities is about unity. 
\bf b:  \rm The mm-wave light curves from MJD$=54593.85$ to $54594.03$. There is an abrupt increase of flux density at MJD$=54593.95$ (black arrow). Within 5 minute, the flux densities at both frequencies became four times or more higher  compared with the value of the previous scan.  
}
\label{Fig4}
\end{center}
\end{figure}
\clearpage
\begin{figure}[thp]
\addtocounter{figure}{-1}	
\begin{center}
\includegraphics[width=14cm]{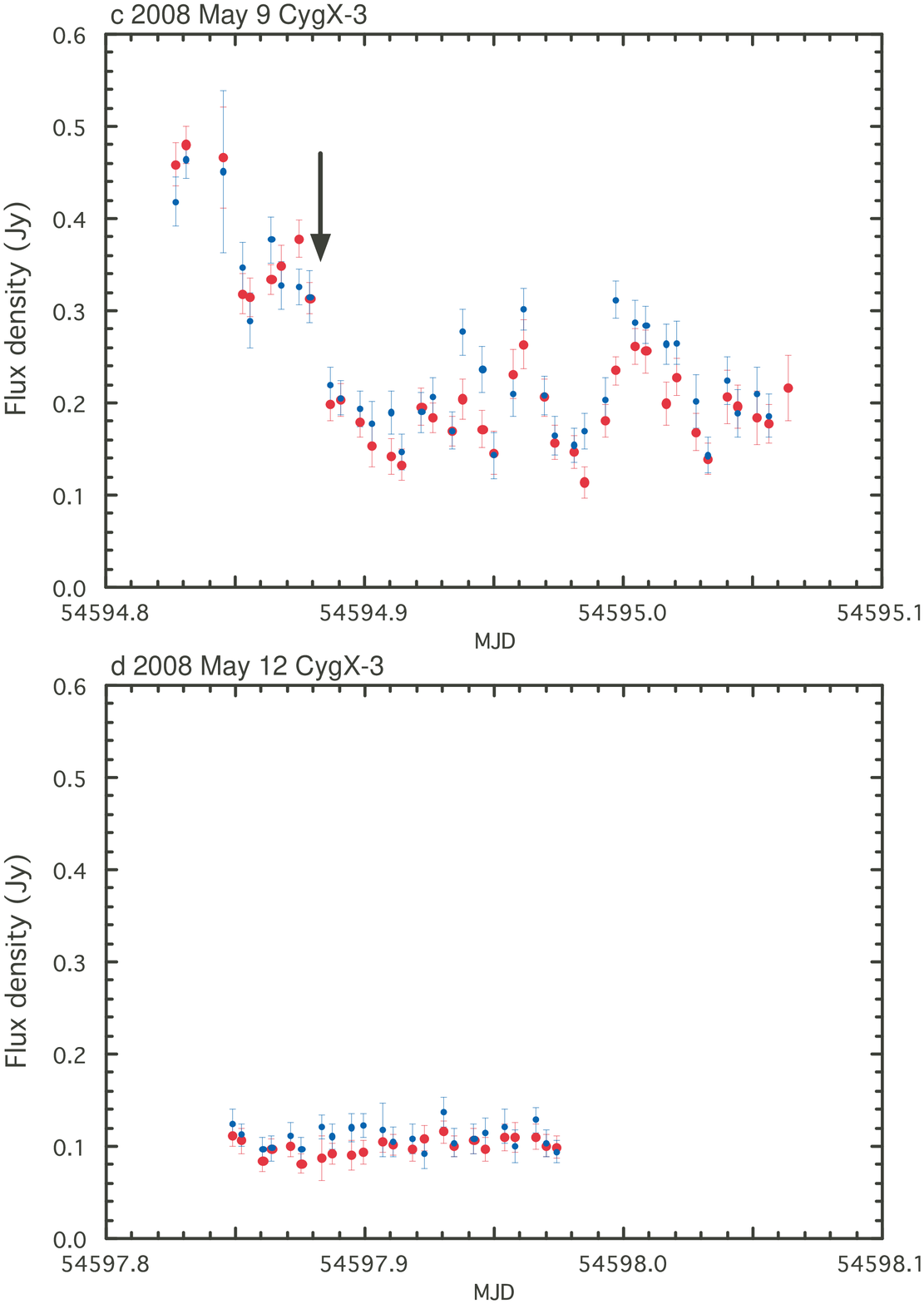}
\caption{\bf\textit{Continued} \bf c:  \rm The mm-wave light curves from MJD$=54594.82$ to $54595.07$. There is probably a similar jump in flux density with opposite-sense at MJD$=54594.88$ (black arrow).  \bf d:  \rm The mm-wave light curves from MJD$=54597.85$ to $54597.98$. They are similar to those of low activity phase.~     }
\label{Fig4b}
\end{center}
\end{figure}
\clearpage

\begin{figure}[thp]
\begin{center}
\includegraphics[width=16cm ]{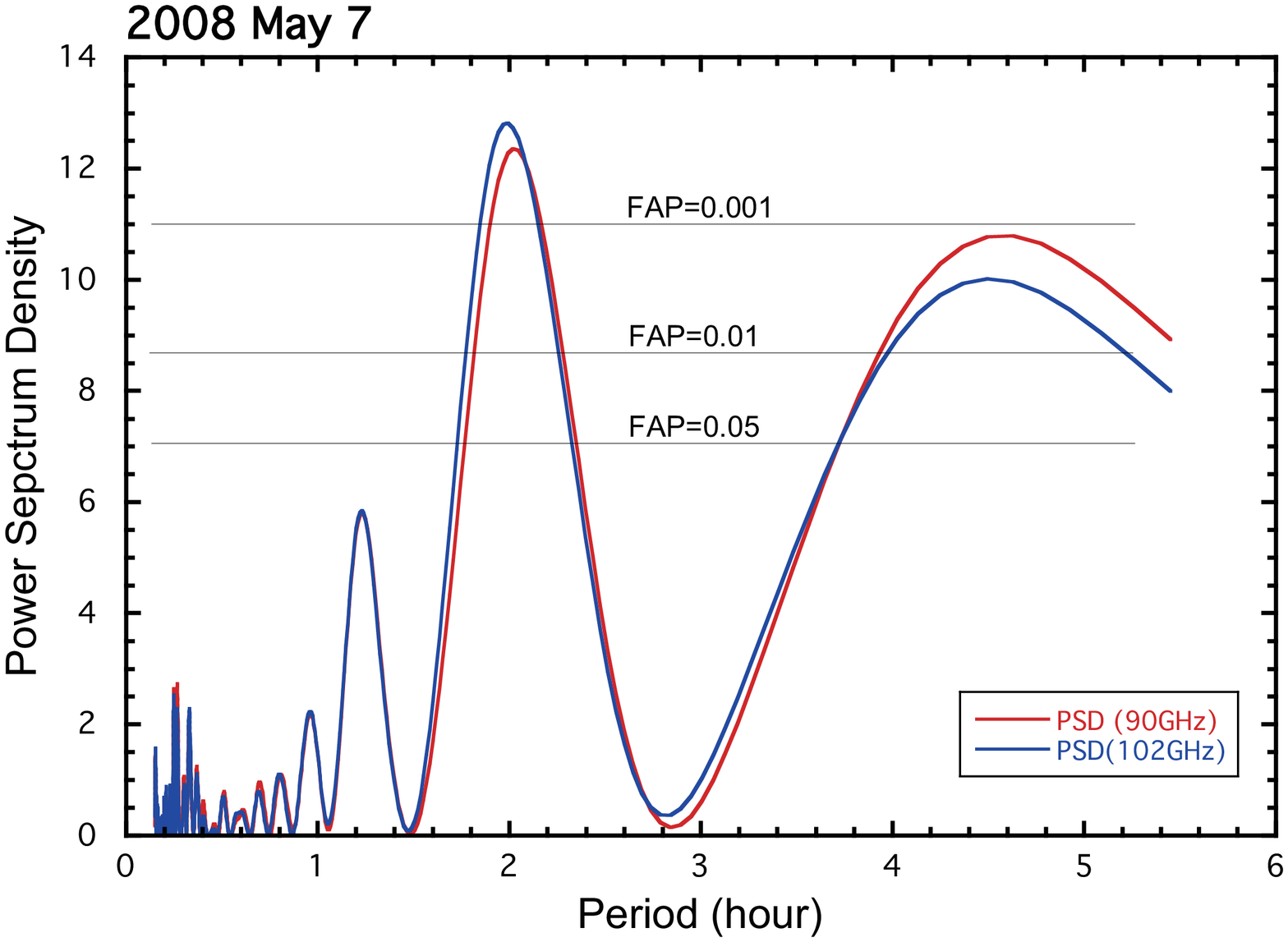}
\caption{Lomb-Scargle periodogram of Cyg X-3 at $\nu=90$ and $102$ GHz  from MJD$=54592.83$ to $54593.07$. The horizontal lines show the false alarm probability; $FAP(>PSD)=1-(1-\exp(-PSD))^M$ (see Eq. 13.8.7 in Press et al. 2007), where $M \sim -6.362+1.193\times N+0.00098\times N^2$ for $N$ data points (see Eq. 13 in Horne and Baliunas 1986).
In this case, the representative values are $FAP(7.098)= 0.050, 
FAP(8.727)= 0.010,$ and 
$FAP(11.034)= 0.001$. There is a peak  with the significance level better than $FAP= 0.001$ at a period of 2 hours.  }
\label{Fig.3 }
\end{center}
\end{figure}


\section{Discussion} 
\subsection{Search for Periodic Variation of Cyg X-3} 
 For another microquasar, GRS~1915+105, a change with quasi-period of 20 minutes has been
detected at 15 and 220 GHz (Fender\&Pooley, 2000). We search any periodic change for the data of MJD$= 54593$ (see Fig 4a), which is free from an abrupt jump.  The Lomb-Scargle method, which is Fourier analysis for non-equal time-interval data, has been applied for the data.  Fig.5 shows the resultant Lomb-Scargle periodogram for the data  from MJD$=54592.83$ to $54593.07$  at $\nu=90$ and $102$ GHz.  The horizontal lines indicate the significance levels.  They are given by the false alarm probability; 
\begin{equation}
\label{ }
FAP(>PSD)=1-(1-\exp(-PSD))^M\end{equation} (see Eq. 13.8.7 in Press et al. 2007), where 
$$M \sim -6.362+1.193\times N+0.00098\times N^2$$ for $N$ data points (see Eq. 13 in Horne and Baliunas 1986).
In this case, the number of data points is $N=55$. The representative values of $FAP$ are  $FAP(7.098)= 0.050, 
FAP(8.727)= 0.010,$ and 
$FAP(11.034)= 0.001$. 
There is a peak with a period of 2 hours in the data.  This has the significance level better than $FAP= 0.001$ although it is not identified as an obvious periodic change in the light curve of  MJD$= 54593$. This period is not consistent with  the orbital period of Cyg X-3, 4.8 hours.  
The origin of the peak is not clear currently.

\subsection{Brightness Temperature of Cyg X-3 at 100 GHz} 
The brightness temperature of the emitting region, $T_\mathrm{B}$, is given by
\begin{equation}
\label{ }
T_\mathrm{B}=\frac{c^2}{2k_B\nu^2}S_{\nu}\Bigl(\frac{d}{r}\Bigr)^2,
\end{equation}
where $S_{\nu}$ is the flux density,  $\nu$ is the observing frequency,
$d$ is the distance of Cyg X-3, and $r$ is the radius of the emitting
region derived from the event in Fig. 4-b. This formula is identical to Equation 1 of Ogley et al. (2001). This  brightness temperature is larger than
$T_\mathrm{B}\ge1.2\times10^{11}$ K assuming that Cyg X-3 is located at a distance of $d=9$ kpc, the observing frequency is $\nu = 100$ GHz, and the flux density is $S_{\nu}= 0.5$ Jy. 

This brightness temperature based on the abrupt intensity jump is as high as one tenth of "inverse-Compton limit" for AGN, $T_\mathrm{B} \le 10^{12}$ K. The ratio of
inverse-Compton scattering loss, $\Bigl(\frac{dE}{dt}\Bigr)_C$, and synchrotron radiation loss,  $\Bigl(\frac{dE}{dt}\Bigr)_S$, of the relativistic electrons is given by
\begin{equation}
\label{ }
R=\frac{\Bigl(\frac{dE}{dt}\Bigr)_C}{\Bigl(\frac{dE}{dt}\Bigr)_S}
=\frac{u_{rad}}{u_{mag}},
\end{equation}
 where $u_{rad}$ and $u_{mag}$ are radiation energy density and magnetic energy density, respectively (see Eq. 5.60 in Pacholczyk 1970), and $H$ is the magnetic field strength of Cyg X-3.   Here, we assume 
$$
u_{rad}\simeq \frac{\nu S_{\nu}}{c}\Bigl(\frac{d}{r}\Bigr)^2, 
$$
because the observed spectral index is flat even at $\nu\sim100$ GHz. 
 , and we also assume
$$
  u_{mag}= \frac{H^2}{8\pi}.
$$
Then
\begin{equation}
\label{ }
R\simeq \frac{3\times 10^{-2}}{H^2}.
\end{equation}
 If inverse-Compton scattering loss dominates synchrotron radiation loss, life time of the relativistic electron 
would be short so that the emission would be invisible.
So we assume that the ratio is $R\le1$. In the case, the magnetic field is estimated to be $H\ge 0.2$ Gauss.


\section{Conclusions} 
 We performed millimeter wave observations of Cyg X-3 experiencing giant outbursts of the 2008 April-May outburst with one of the finest time resolutions. 
We found the series of short-lived flares with amplitude 1-2 Jy. They have flat spectra around 100 GHz.  
We also found an abrupt and large amplitude flux density change with e-folding time of 3.6 
minutes or shorter in the millimeter light curve. This requires a small-scale structure of 0.4 AU in the jet of Cyg X-3.  The brightness temperature is estimated to be $T_B\sim 1 \times 10^{11}$ K. 
If so, the structure will be able to be imaged by the up-coming mm-wave space-VLBI, VSOP2.

\bigskip 
The authors thank anonymous referee for informative comments. They also thank the members of NMA group of Nobeyama Radio Observatory for support in the observations. 

\end{document}